\documentclass[aps,prl,twocolumn,superscriptaddress,showpacs]{revtex4}
\usepackage{amsmath}
\usepackage{graphicx,epsfig}
\usepackage{subfigure}

\newcommand{\be}{\begin{equation}}
\newcommand{\ee}{\end{equation}}
\newcommand{\nn}{\nonumber\\}
\newcommand{\ij}{\langle ij \rangle}

\newcommand{\ba}{\begin{eqnarray}}
\newcommand{\ea}{\end{eqnarray}}

\begin{document}

\title{Valence-bond-solid order in antiferromagnets with spin-lattice coupling}
\author{Chenglong Jia}
\affiliation{Department of Physics and Institute for
Basic Science Research, \\
Sungkyunkwan University, Suwon 440-746, Korea}
\author{Jung Hoon Han}
\email[Electronic address:~]{hanjh@skku.edu} \affiliation{Department
of Physics and Institute for
Basic Science Research, \\
Sungkyunkwan University, Suwon 440-746, Korea} \affiliation{CSCMR,
Seoul National University, Seoul 151-747, Korea}
\date{\today}

\begin{abstract}
We propose that a valence-bond-solid (VBS) order can be stabilized
in certain two-dimensional antiferromagnets due to spin-lattice
coupling. In contrast to the VBS state of the
Affleck-Kennedy-Lieb-Tesaki (AKLT) type, the spin-lattice
coupling-induced VBS state can occur when $2S$ is not an integer
multiple of the coordination number $z$. $S=2$ spins on the
triangular lattice with $z=6$ is discussed as an example. Within the
Schwinger boson mean-field theory the ground state is derived as the
direct product of states, one of which represents the conventional
long-range ordered spins, and the other given by the
$\sqrt{3}\times\sqrt{3}$ modulation of the valence bond amplitudes.
Excitation spectrum for the modulated valence bond state is worked out within the
single-mode approximation. The spectrum offers a new collective
mode, distinct from the spin wave excitations of the magnetically
ordered ground state, and observable in neutron scattering.
\end{abstract}

\maketitle

Spin-Peierls phenomenon refers to the dimerization of
antiferromagnetic $S=1/2$ spins on a linear chain accomapanied by
spontaneous lattice distortion\cite{pytte,CuGeO3}. The physics
underlying the spin-Peierls transition is the extra gain in energy
through the reduction of dimerized bond length. A similar phenomenon
is believed to occur in two or higher dimensions\cite{starykh}.
Another route to obtaining dimerized ground state is by including
interactions beyond the nearest-neighbor Heisenberg exchange, as in
the Majumdar-Ghosh model for one-dimensional antiferromagnetic
chain\cite{MG}.

Generalization of Majumdar-Ghosh idea to two dimensions and to
arbitrary spin $S$, and construction of a general class of exact
ground states to antiferromagnetic spin models were carried out by
Affleck, Kennedy, Lieb, and Tesaki (AKLT)\cite{AKLT}. AKLT's idea,
and the subsequent refinement by Arovas, Auerbach, and Haldane
(AAH)\cite{AAH}, emphasize the close connection between the spin
quantum number $2S$ and the lattice coordination, $z$. A
valence-bond-solid (VBS) state is formed when $2S$ is an integer
multiple of $z$, and every nearest-neighbor bond is covered by
$2S/z$ dimers.

In this paper, we address the case where the two numbers, $2S$ and
$z$, are incommensurate. An example would be $S=2$ antiferromagnet
on a triangular lattice, where the AKLT condition would require
$S=3$. Such a system is realized, for example, in the compounds
RMnO$_3$(R=Lu,Y)\cite{RMnO3-exp,pisarev,park}. In these systems,
each magnetic Mn ion carries spin $S=2$.  At most four valence bonds
can be formed from each site, leaving the other two bonds
``unfulfilled". Long-range magnetic order has been found in these
compounds. Furthermore, there is a strong indication of the movement
of the Mn position concurrent with the magnetic transition, pointing
to the existence of spin-lattice coupling in these
materials\cite{park}.

Attempts to understand the effect of spin-lattice coupling for the
triangular network of Heisenberg spin for large $S$ has been given
in Ref. \onlinecite{jia}. It was shown that the mean-field theory
predicts lack of lattice displacement as the magnetic order builds
up, in contrast to the experimental observation. Other types of
order, such as the valence bond order, had not been examined in
detail. It is shown here that a type of VBS order with only four of
the six bonds of the triangular lattice being filled by singlets can
be stabilized through spin-lattice interaction and leads to lattice
deformation as in the spin-Peierls systems.

\begin{figure}[b]
\includegraphics[width=6cm]{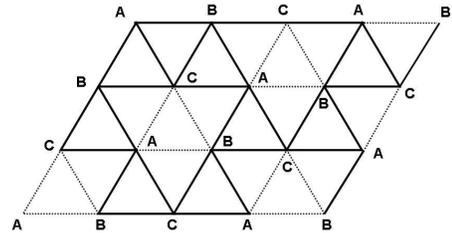}
\caption{{\protect\small Realization of partial VBS order on a
triangular lattice with spin $S=2$. At most four valence bonds can
emanate from each site (thick lines), forming a regular array of
``missing bonds" (dotted triangles). Magneto-elastic coupling can
alter the bond distance depending on the presence/absence of the
valence bond. Three inequivalent sites are denoted A, B, and C.}}
\label{our-idea}
\end{figure}

Our ideas are illustrated in Fig. \ref{our-idea} for $S=2$ and $z=6$
(triangular lattice). Such states are energetically unfavorable in comparison to the
magnetically ordered state which makes use of the exchange energy
for all six neighbors. On the other hand, the spin-lattice coupling
can reduce the distance for a pair of sites forming a valence bond,
and increase it for sites lacking a bond. The gain in the exchange
energy achieved in this way will overcome the loss from the two
missing bonds for a sufficiently large coupling strength, and the
VBS state is energetically favored over the magnetically ordered
state.

We provide a quantitative justification of our claim by analyzing
the lattice-coupled spin Hamiltonian,
\begin{equation}
H = \sum_{\langle ij \rangle} J_{ij} S_i \cdot S_j +
{\frac{K}{2}}\sum_i u_i^2 , \label{our-model}
\end{equation}
taking $J_{ij} = J_0 (d^0_{ij}/d_{ij})^\gamma$ with the equilibrium
and actual ionic separations given by $d_{ij}^0$ and $d_{ij}$,
respectively. For transition metal ions $\gamma$ is a
rather large number, ranging between 6-14\cite{harrison}.
Displacement of the $i$-th site is given by
$u_i$, which costs an elastic energy of $(K/2)u_i^2$. For ease of
analysis, we Taylor-expand $J_{ij}$ in small displacements as
$J_{ij} = J_0 [ 1-\gamma \hat{e}_{ji} \cdot (u_j - u_i ) ]$, taking
$d^0_{ij} =1$, and $\hat{e}_{ji}=(r_j - r_i)/|r_j - r_i|$ is the
unit vector connecting the equilibrium ionic sites, $r_i$ and $r_j$.

In the following we re-define $u_i \rightarrow u_i /\sqrt{K}$ and
absorb $\gamma/\sqrt{K}$ as $\alpha$, while the overall energy scale
$J_0$ is normalized to one. Because of the large value of $\gamma$,
a small displacement can lead to a substantial difference in the
exchange energy. Only the adiabatic limit of the phonon dynamics is
considered as we are ignoring the kinetic energy. The model is
defined for the two-dimensional triangular lattice with the
nearest-neighbor exchange interaction $\langle ij \rangle$ only. A
similar model on a linear chain and on a square lattice has been
studied earlier\cite{mila}. We look for the ground state of the
above Hamiltonian within the Schwinger boson mean-field theory
(SBMFT)\cite{AA}.

One re-writes the above Hamiltonian using a pair of Schwinger bosons
$(b_{i1},b_{i2})$ defined at each site,
\begin{equation}
S_i \cdot S_j \rightarrow - {\frac{1}{2}} A^+_{ij} A_{ij},
\end{equation}
where $A_{ij}=b_{i1}b_{j2}-b_{i2}b_{j1}$, and decouple it by
introducing the order parameter $Q_{ij} = \langle A_{ij}\rangle = -
Q_{ji}$. Non-zero $|Q_{ij}|$ implies  finite amplitude of a singlet
bond for the $\langle ij \rangle$ link. The resulting mean-field
Hamiltonian is
\begin{eqnarray}
H_{MF}&=& - {\frac{1}{2}}\sum_i \sum_{j\in i}\overline{\Delta}_{ij}
b_{i1}b_{j2} -{\frac{1}{2}}\sum_i \sum_{j\in i}\Delta_{ij}
b^{+}_{j2}b^{+}_{i1}  \notag \\
&&+\sum_i \lambda_i (b^{+}_{i1}b_{i1}+b^{+}_{i2}b_{i2} -2 S) +
{\frac{1}{2}}\sum_i u_i^2, \label{H_mf}
\end{eqnarray}
where we have introduced $\Delta_{ij} \equiv J_{ij} Q_{ij} $, and $j
\in i$ to indicate all nearest neighbors of $i$. Lagrange multipliers
$\lambda_i$ are fixed by the constraint that
$\partial \langle H_{MF} \rangle /\partial \lambda_i = 0$ which also
amounts to requiring $\langle b^{+}_{i1}b_{i1}+b^{+}_{i2}b_{i2}
\rangle =2S$ at each site. Energy minimization
requires that $u_i$ obeys\cite{jia,mila}
\begin{equation}
u_i =\alpha \sum_{j\in i} \hat{e}_{ij} \langle S_i \cdot S_j \rangle = -{\frac{%
\alpha}{2}}\sum_{j\in i} \hat{e}_{ij} |Q_{ij} |^2 .  \label{u-to-S}
\end{equation}

We proceed to solve the mean-field Hamiltonian assuming  (i)
homogeneous pairing and (ii) inhomogeneous pairing of the type
described in Fig. \ref{our-idea}, and compare their energies.

For the homogeneous case, the singlet pair amplitudes are taken as $
Q_{i,i+e_1}=Q_{i,i+e_2}=Q_{i,i+e_3}=iQ$ and $\lambda_i =\lambda$.
The three unit vectors are defined by $\hat{e}_1 = (1,0)$, $\hat{e}_2 =
(-1/2,\sqrt{3}/2)$, $\hat{e}_3 = (-1/2,-\sqrt{3}/2)$. Passing to momentum space,
the Hamiltonian becomes

\begin{equation}
H_{MF} = \sum_k (
\begin{array}{cc}
b^{+}_{k1} & b_{-k2}%
\end{array}%
) \left(%
\begin{array}{cc}
\lambda & \Delta_k \\
\Delta_k & \lambda%
\end{array}%
\right) \left(%
\begin{array}{c}
b_{k1} \\
b^{+}_{-k2}%
\end{array}
\right) ,  \label{H}
\end{equation}
where $\Delta_k = Q \sum_{\alpha=1}^3 \sin ( k \cdot \hat{e}_\alpha ) $.
The equation of motion for the boson operators can be derived from
Eq. (\ref{H}) and utilizing the boson commutation algebra,

\begin{equation}
E_{k}\left(
\begin{array}{c}
u_{k} \\
v_{k}%
\end{array}%
\right) =\left(
\begin{array}{cc}
\lambda & \Delta _{k} \\
-\Delta _{k} & -\lambda%
\end{array}%
\right) \left(
\begin{array}{c}
u_{k} \\
v_{k}%
\end{array}%
\right) ,
\end{equation}
which has a pair of eigenvalues $\pm E_k$, $E_{k}=\sqrt{\lambda
^{2}-\Delta _{k}^{2}}$.
Boson occupation numbers $n_i =\langle
b_{i1}^{+}b_{i1}+b_{i2}^{+}b_{i2} \rangle $, and the pairing
amplitudes are worked out
\ba  n_i +1 = {\frac{1}{N_k }}\sum_{k}{\frac{\lambda }{E_{k}}} =2S
+1, \nn  Q = {\frac{1}{3N_k }}\sum_{k}\sum_{\alpha
=1}^{3}{\frac{\Delta _{k}}{E_{k}}}\sin (k\cdot e_{\alpha }).
\label{SC-for-uniform} \ea
The $k$-sum is restricted to the first Brillouin zone of the
triangular lattice, and $N_k$ are the number of distinct $k$-points.
Upon solving Eq. (\ref{SC-for-uniform}) we obtain $Q=\sqrt{3}%
S+0.1161$ and the ground-state energy per site
\begin{equation}
E^h_{g.s.}=-{\frac{3}{2}}Q^{2}=-{\frac{3}{2}}(\sqrt{3}S+0.1161)^{2}.
\label{E-for-uniform}
\end{equation}
There is no energy gain from magneto-elastic coupling since lattice
displacement is zero for the uniform bond amplitudes.

Next, we consider the situation where two distinct bond amplitudes,
$Q_1$ and $Q_2$, corresponding to dotted and thick lines in Fig.
\ref{our-idea}, exist. Non-zero lattice displacement develops with
magnitude $|u_i | =(\sqrt{3}\alpha/2) ( Q_2^2 -Q_1^2 )$, and the
directions consistent with the uniform expansion of those triangles
in Fig. \ref{our-idea} lacking a valence bond. Exchange energy
$J_{ij}$ is similarly modulated, $ J_1 = 1-(3\alpha^2 /2) ( Q_2^2 -
Q_1^2 )$, $J_2 = 1 + (3\alpha^2 /4) (Q_2^2 -Q_1^2 )$, and finally,
$\Delta_1 = J_1 Q_1$ and $\Delta_2 = J_2 Q_2$.

Corresponding equation of motion of the operators can be worked out.
Six-dimensional eigenvectors $\psi_k = ( u_{Ak} ~ u_{Bk} ~ u_{Ck} ~ v_{Ak} ~
v_{Bk} ~ v_{Ck} )^T$ satisfy an equation of motion $E_k \psi_k = M_k \psi_k$
where

\begin{equation}
M_k = \left(
\begin{array}{cccccc}
\lambda & 0 & 0 & 0 & -i\eta_{1k} & i\overline{\eta}_{3k} \\
0 & \lambda & 0 & i\overline{\eta}_{1k} & 0 & -i\eta_{2k} \\
0 & 0 & \lambda & -i\eta_{3k} & i\overline{\eta}_{2k} & 0 \\
0 & i\eta_{1k} & -i\overline{\eta}_{3k} & -\lambda & 0 & 0 \\
-i\overline{\eta}_{1k} & 0 & i\eta_{2k} & 0 & -\lambda & 0 \\
i\eta_{3k} & -i\overline{\eta}_{2k} & 0 & 0 & 0 & -\lambda%
\end{array}%
\right), \label{EoM}
\end{equation}
and
\begin{eqnarray}
2\eta_{1k} &=& \Delta_1 e^{i k\cdot e_1} + \Delta_2 e^{i k\cdot e_2 } +
\Delta_2 e^{i k \cdot e_3}  \notag \\
2\eta_{2k}&=& \Delta_2 e^{i k\cdot e_1} + \Delta_1 e^{i k\cdot e_2 } +
\Delta_2 e^{i k \cdot e_3}  \notag \\
2\eta_{3k}&=& \Delta_2e^{i k\cdot e_1} + \Delta_2 e^{i k\cdot e_2 } +
\Delta_1 e^{i k \cdot e_3} .
\end{eqnarray}
Each eigenvector is normalized to obey $|u_{Ak} |^2 + |u_{Bk} |^2 + |u_{Ck}
|^2 - |v_{Ak} |^2 -|v_{Bk} |^2 - |v_{Ck} |^2 = 1$.

While the problem is not tractable analytically, it can be shown
that the (extreme) limit of $\Delta_1 = 0$ yields an analytical
answer. We call it the ``kagome limit", where the non-zero
$\Delta$'s  form a network topologically equivalent to the kagome
lattice. Using the linear Taylor expansion of exchange energy $J_1$
and $J_2$, the ground state energy is given by
\be E^m_{g.s.} = - {1\over 2} Q_1^2  -Q_2^2  - {3\over 8} \alpha^2
(Q_2^2 - Q_1^2 ) ^2, \label{E-for-nonuniform}\ee
for the lattice-distorted situation. In the kagome limit the
self-consistency equations give $Q_1= \sqrt{3}S - 0.2128$, and $Q_2
= \sqrt{3} S + 0.1823$. Finally, in order to make $J_1 = 0$,
$\alpha$ must be equal to $0.4946$. In comparing Eq.
(\ref{E-for-uniform}) with Eq. (\ref{E-for-nonuniform}) with $S=2$,
the state with modulated bond is found to have a lower energy.

The stability of the bond-modulated state can be argued on a more
general basis. Restoring the various energy scales, the ground state
energies of homogeneous and modulated bonds are given by

\be E^h_{g.s.} = -{3\over 2}J_0 Q^2 , ~~ E^m_{g.s.} = -{1\over 2} J_1 Q_1^2 -
J_2 Q_2^2 + {K\over 2} u^2 .\ee
As our analysis in the kagome limit showed, $Q$ depends slowly on
the amount of distortion $u$, and all three $Q$ values are given by
$Q\approx Q_1 \approx Q_2 \approx \sqrt{3}S$. Much more dramatic
dependence on $u$ arises from $J_1$ and $J_2$, which reads
$J_{1}=J_0 /(1+\sqrt{3}u)^\gamma$ and $J_2 = J_0
/(1-\sqrt{3}u/2)^\gamma$ assuming small $u$. Using $Q =Q_1 = Q_2=
\sqrt{3}S = 2\sqrt{3}$, the energy difference
$E^h_{g.s.}-E^m_{g.s.}$ is shown to be a monotone decreasing
function of $u$ for $K/J_0 \le 27 \gamma (\gamma+1)$, indicating a
lattice instability. The quartic term in the potential energy for
$u$ not included in Eq. (\ref{our-model}) will restore the stability
at a finite $u$ and favor the lattice reconstruction indicated in
Fig. \ref{our-idea}.

The mean-field ground state features Bose condensation for both
uniform and non-uniform states which are the hallmarks of magnetic
order. For $k=k^*$ at which the Bose condensation occurs, the ground
state is described by a form $|\mathrm{LRO}\rangle =
(\gamma^+_{k^*})^{N_{k^*}} |0\rangle$ ($N_{k^*}$ = number of bosons
in the condensed state). For other $k$ values the ground state is
defined by $\gamma_k | \mathrm{MF} \rangle = 0$, where $\gamma_k$
are the quasiparticle operators diagonalizing the mean-field
Hamiltonian, Eq. (\ref{H_mf}). The uncondensed part, written in real
space, is of the general form,
$ |\mathrm{MF}\rangle = \exp \left\{ {\frac{1}{2}}\sum_{ij} u_{ij}
(b^{+}_{i1}b^{+}_{j2}-b^{+}_{j1}b^{+}_{i2}) \right\} |0\rangle,
\label{MF-state} $
representing a gapped spin liquid\cite{AA,Book}. The sensitivity of the ground
state on a particular spin quantum number $S$ arises when we make a
projection to a fixed $S$: $|\mathrm{VBS}\rangle = P_S
|\mathrm{MF}\rangle$. \textit{The full ground state is a direct
product of the condensed and the uncondensed parts},
$|\mathrm{GS}\rangle = |\mathrm{LRO}\rangle \otimes
|\mathrm{VBS}\rangle$.

With $S=2$, the projected ground state consists of a large number of
valence bond configurations all sharing the property that each site
has four valence bonds emanating form it. Among those states, the
VBS state depicted in Fig. \ref{our-idea} is the one most consistent
with the kagome limit. Such a state, in turn, is described by the
AKLT state:
\begin{equation}
|\mathrm{AKLT}\rangle =\prod_{\ij '}
(b_{i1}^{+}b_{j2}^{+}-b_{i2}^{+}b_{j1}^{+})|0\rangle. \label{AKLT}
\end{equation}
$\ij' $ runs over the four valence bonds depicted in Fig. \ref{our-idea}.

Although  the equivalence of  $|\mathrm{AKLT} \rangle$ with  $|
\mathrm{VBS} \rangle_{S=2}$ cannot be established rigorously, it is
still reasonable to expect that the physical properties embodied in
the $|\mathrm{AKLT}\rangle$ state will reflect those of the $|
\mathrm{VBS} \rangle_{S=2}$ \cite{Raykin}.

\begin{figure}[t]
\includegraphics[width=5cm]{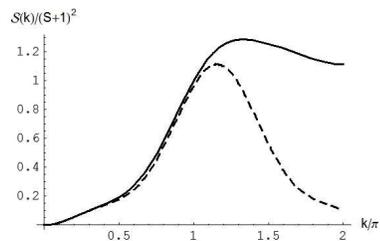}
\caption{{\protect\small Static structure factor ${\cal
S}(k)=\langle S_{-k}\cdot S_k \rangle$ for the
$|\mathrm{AKLT}\rangle$ state in Eq. (\ref{AKLT}). Solid line is
${\cal S}(k,0)$, and the dashed line is ${\cal S}(0, k)$. MC
calculations were performed on  $N=90\times90$ punctured triangular
lattice. }} \label{SSFactor}
\end{figure}

\begin{figure*}
\centering \mbox{ \subfigure{\includegraphics[width=6cm]{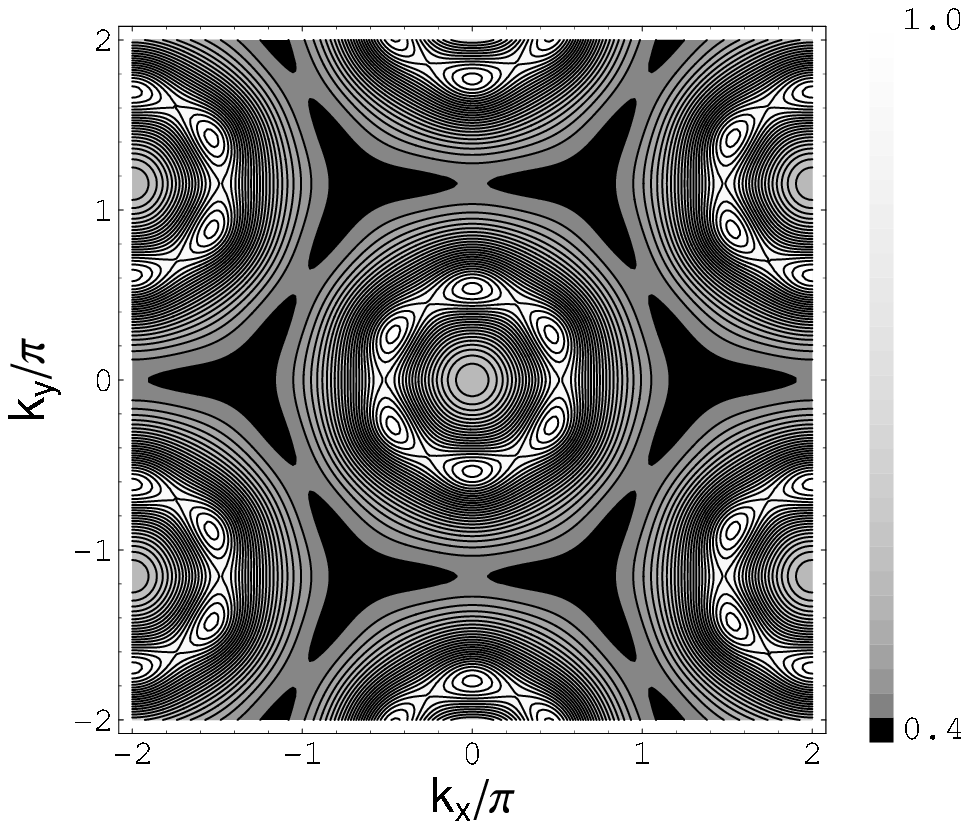}}
\subfigure{\includegraphics[width=6cm]{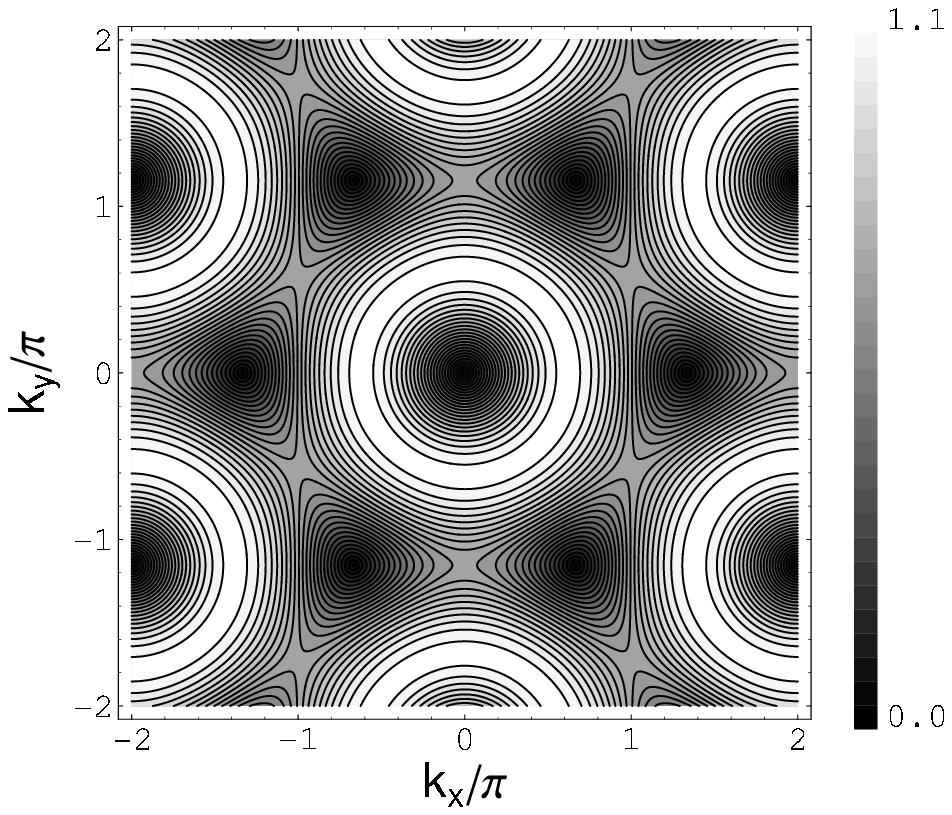}}
\subfigure{\includegraphics[width=5cm,height=5cm]{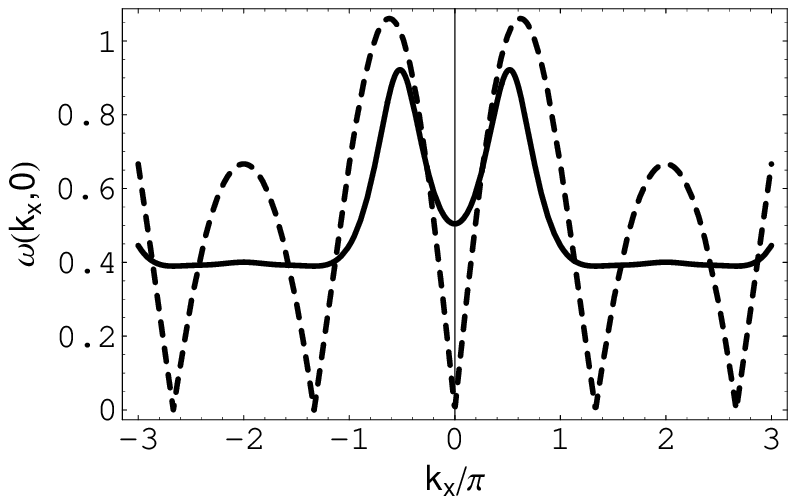}}}
\caption{{\protect\small  Contour of the energy spectrum for (a) VBS
state given in Eq. (\ref{AKLT}),  and (b) spin waves in triangular
lattice. Both plots are scaled by $6J$. (c) Disperson along $(k,0)$
direction for VBS (solid) and spin wave (dashed). }} \label{Energy}
\end{figure*}

The quasiparticle spectrum obtained from diagonalizing Eq.
(\ref{EoM}) reduces to the familiar spin wave modes\cite{mattson}.
The excitation energies for the AKLT ground state given above can be
worked out in a manner parallel to the calculation for the gapped
spin liquid in one-dimensional, $S=1$ chain\cite{AAH}. We adopt the
method of single mode approximation (SMA), and write the excitation 
energy $\Delta(k)$:
\ba &&\Delta (k) =\frac{\langle \lbrack S_{-k}\cdot ,[H,
S_{k}]\rangle }{2 \langle S_{-k} \cdot S_k \rangle } \nn && =2z'
B_{1}\times {\frac{1-(1/3)\sum_{\alpha =1}^{3}\cos [k\cdot e_{\alpha
}] }{(1/N)\sum_{i,j}\langle S_i \cdot S_j \rangle e^{i k\cdot (r_j -
r_i ) }}}. \label{new-excitation} \ea
The averages $\langle \cdots \rangle$ are to be taken with respect
to $|\mathrm{AKLT} \rangle$ state given in Eq. (\ref{AKLT}). The
nearest-neighbor spin-spin correlation $-B_{1}$, as well as all
other averages $\langle S_i \cdot S_j \rangle$ can be evaluated
using the standard Monte Carlo (MC) method\cite{Book,MC}. The
effective coordination number on the kagome  lattice is $z'=4$; the
number of sites of the triangular lattice is $N$.

The structure factor ${\cal S}(k )\equiv \langle S_{-k} \cdot S_k
\rangle $ vanishes in the $k = 0$ limit due to the singlet character
of the AKLT state. The next term in the small-$k$ expansion gives
${\cal S}(k)\approx \Gamma k^{2}$, where $-\Gamma
=(1/N)\sum_{i,j}(r_{j}-r_{i})^{2}\langle S_{i}\cdot S_{j}\rangle $.
The numerator in Eq. (\ref{new-excitation}) also vanishes as
$k^{2}$, giving the $k=0$ gap of magnitude $\Delta (k=0) =8J \times
( B_{1} / \Gamma )$.

The AKLT state  has only short-range order as manifested by an
exponential decay of  the spin-spin correlation, $|\langle S_i
\cdot S_j \rangle |\sim e^{-1.467 |i-j|} $, in our MC simulation.
In Fig. \ref{SSFactor} we present results for the static structure
factor. The $k =0$ gap obtained from ${\cal S}(k)$ is $\Delta
(0)\simeq  0.5 \times (6J)$.

Figure \ref{Energy} displays excitation spectrum $\Delta(k)$ for
$|\mathrm{VBS}\rangle $ according to Eq. (\ref{new-excitation}), and
the standard spin wave in the uniform $|\mathrm{LRO}\rangle$. The
new spectrum $\Delta(k)$ is characterized by a broad minimum (dark
region) that exist around the zone boundary of the Brillouin zone,
and in principle should be observable in the neutron scattering. The
spin waves and the new excitation mode we find reflect the two kinds
of broke symmetries found in the ground state. Interaction between
the two modes should exist, which requires going beyond the
quadratic Hamiltonian (\ref{H_mf}) to include fluctuation terms.

In summary, we propose that a VBS state can be stabilized in
two-dimensional antiferromagnetic spins with a mismatch of $2S$ and
$z$, when a sufficiently strong coupling to the lattice deformation
exists. As an example, we analyze $S=2$, $z=6$ (triangular) model in
the Schwinger boson mean-field theory and show that the energy is
lower for the lattice-deformed state. Corresponding gapped spin
excitation energies are calculated within the single-mode
approximation.

\begin{acknowledgments}
We thank Sang-Wook Cheong and Je-Geun Park for insightful
discussions. H.J.H. is supported by Korea Research Foundation Grant
(KRF-2004-015-C00181).
\end{acknowledgments}

\end{document}